\documentstyle[epsfig]{aipproc}

\begin{document}
\title{Gamma Ray Pulsars:  Observations}

\author{David J. Thompson}
\address{Laboratory for High Energy Astrophysics\\NASA Goddard Space Flight Center\\Greenbelt, Maryland 20771 USA}

\maketitle

\begin{abstract}
High-energy gamma rays are a valuable tool for studying particle acceleration and radiation in the magnetospheres of energetic pulsars.  The six or more pulsars seen by CGRO/EGRET show that: the light curves usually have double-peak structures (suggesting a broad cone of emission); gamma rays are frequently the dominant component of the radiated power; and all the spectra show evidence of a high-energy turnover.  Unless a new pulsed component appears at higher energies, progress in gamma-ray pulsar studies will be greatest in the 1-20 GeV range.   Ground-based telescopes whose energy ranges extend downward toward 10 GeV should make important measurements of the spectral cutoffs. The Gamma-ray Large Area Space Telescope (GLAST), now in planning for a launch in 2005, will provide a major advance in sensitivity, energy range, and sky coverage. 
\end{abstract}

\section*{Introduction}
%
%
%
%

Pulsars are particularly interesting astrophysical objects because we know so 
much about them, and much of that information is derived from their timing 
properties.     Measurement of the period P and period derivative $\dot{P}$ first identified 
pulsars as compact objects.  Under fairly general assumptions that pulsars are rapidly-
rotating neutron stars with a strong dipole magnetic field, a whole range of physical 
parameters can be estimated from these timing parameters.	Examples include the 
timing age, the spin-down energy loss, the surface magnetic field, and the open field 
line voltage [1].

\section*{The Observed Gamma-Ray Pulsars}
							
The telescopes on the Compton Gamma Ray Observatory identified a number of  gamma-ray pulsars, some with very high confidence and others with less certainty.   There are at least three answers to the question, ``How many gamma-ray pulsars are there?'' 

The first answer is that there are seven or more gamma-ray pulsars. Figure 1 shows their light curves in four energy bands: roughly 0.5 - 2 KeV, 2 - 100 keV, 100 keV - 10 MeV, and above 100 MeV.  All seven of these are positive detections in the gamma-ray band above 100 KeV. The weakest (PSR B1951+32) has a statistical probability of occurring by chance of $\sim$10$^{-9}$.

\begin{figure}[b!] 
\centerline{\epsfig{file=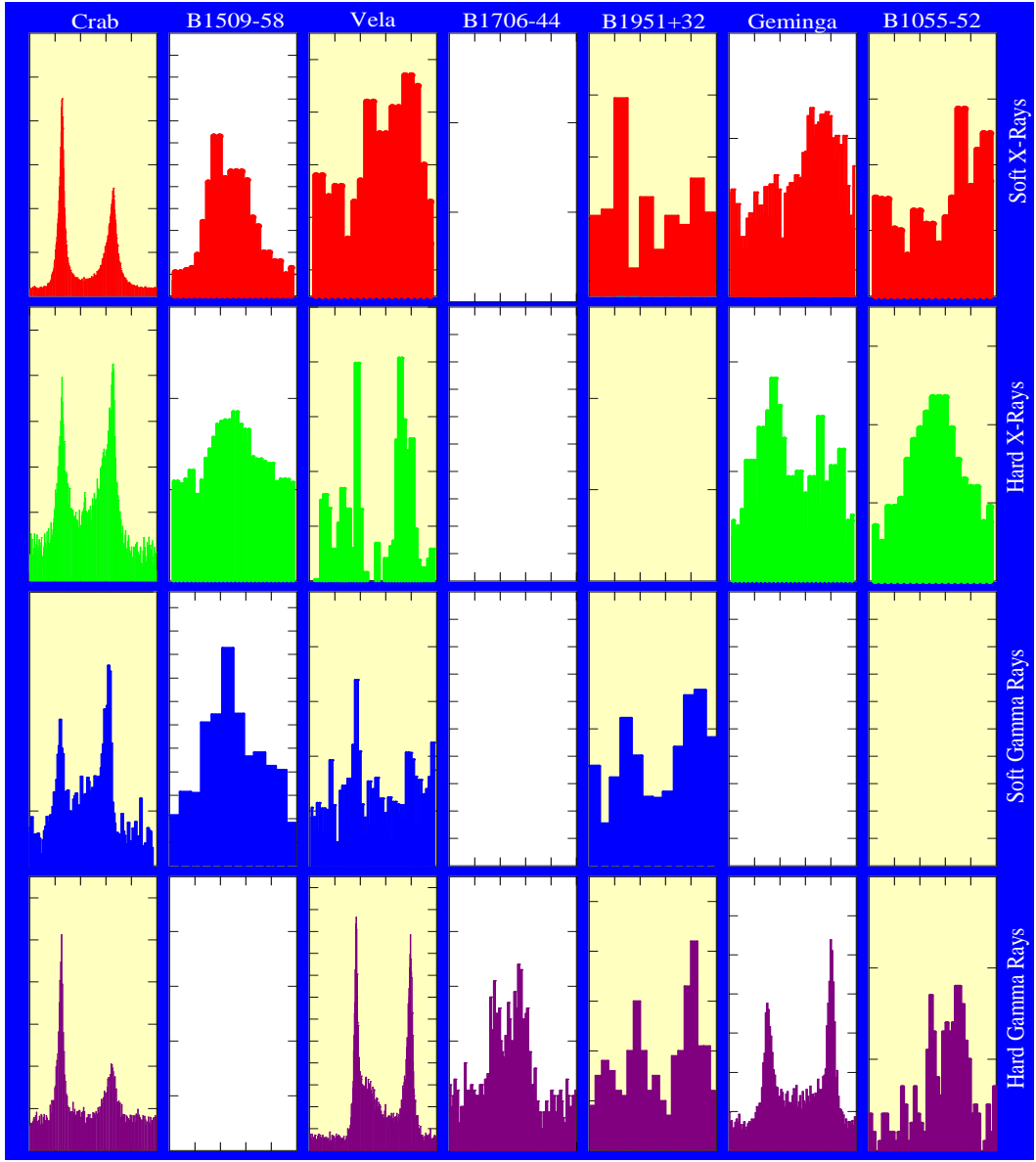,height=6.2in,width=5.5in}}
\vspace{10pt}
\caption{Light curves of seven gamma-ray pulsars in X-rays and gamma rays,  from left to right in order of timing age. Each panel shows one full rotation of the neutron star. 
Ref. Crab: [2-4, this work]; PSR B1509-58: [5-7]; Vela: [8-11]; PSR B1706-44: [12]; 
PSR B1951+32: [13-15]; Geminga: [16, this work]; PSR B1055-52: [17].}
\label{fig1}
\end{figure}

Some important features of these pulsar light curves are:
\begin{itemize}
\item 
 They are not the same at all wavelengths.  Some combination of the geometry and the emission mechanism is energy-dependent.  In soft X-rays, for example, the emission in same cases seems to be thermal, probably from the surface of the neutron star; clearly this is not the case for gamma radiation.

\item Not all seven are seen at the highest energies.  PSR B1509$-$58 is seen only up to 10 MeV, by COMPTEL [7], and not above 100 MeV by EGRET.  For this reason, the answer to  the question, ``How many {\it high-energy} gamma ray pulsars are seen?'' is, ``At least six.''

\item The six seen by EGRET all have a common feature - they show a double peak in their light curves.  Because these high-energy gamma rays are associated with energetic particles, it seems likely that the particle acceleration and interactions are taking place along a large hollow cone or surface.
\end{itemize}

\epsfclipon
\begin{figure}[b!] 
\centerline{\epsfig{file=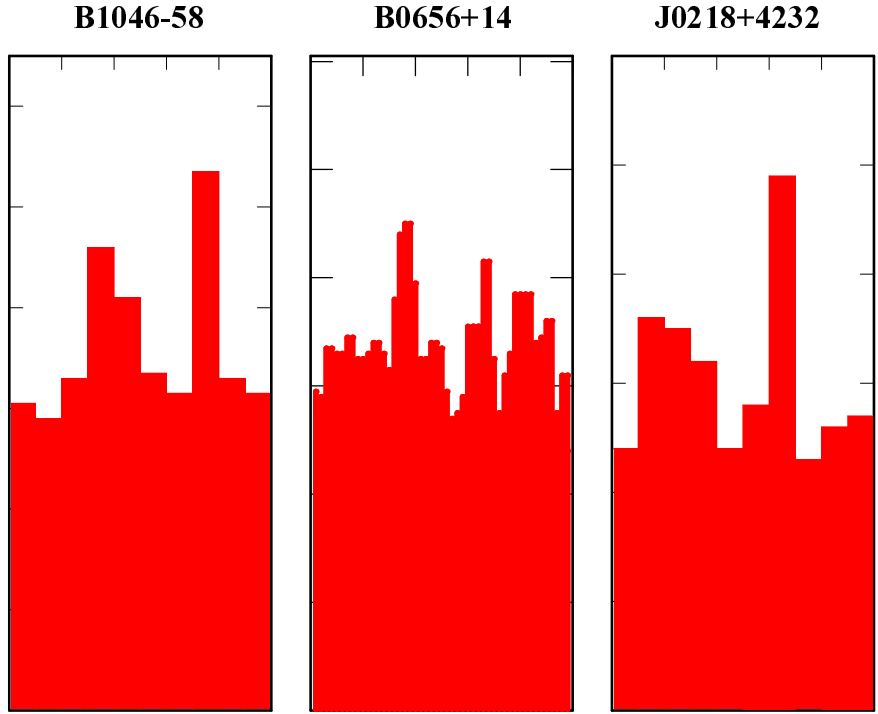,height=4.in}}
\vspace{10pt}
\caption{Light curves of three candidate gamma-ray pulsars. Each panel shows one full rotation of 
the neutron star.  References: PSR B1046-58: [18]; PSR B0656+14: [19]; PSR J0218+4232 [20].}
\label{fig2}
\end{figure}

In addition to the six high-confidence pulsar detections above 100 MeV, three additional radio pulsars may have been seen by EGRET.  Figure 2 shows their light curves (without the zero suppression used in some of the original references).  These three all have statistical probabilities in the 10$^{-4}$ range, or about 5 orders of magnitude less convincing than the weakest of the seven on the previous figure.  These are good candidates, but they are not strong enough to be used as discriminators between models.  These three do imply a third answer to the ``how many'' question - there could be as many as ten known gamma-ray pulsars.

\begin{figure}[b!] 
\centerline{\epsfig{file=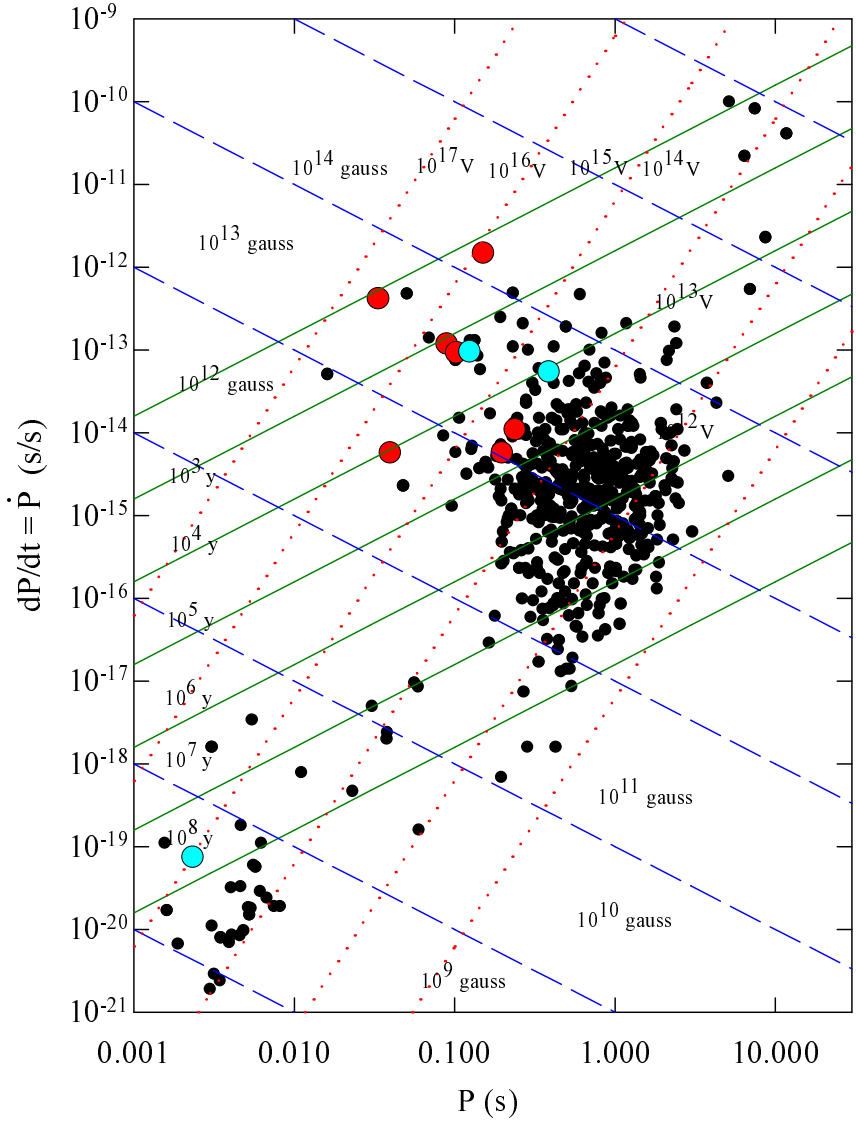,height=6.in,bbllx=20pt,bblly=30pt,bburx=620pt,bbury=780pt,clip=.}}
\vspace{10pt}
\caption{Period v. period derivative for a large sample of pulsars. Small dots: no gamma-ray emission. Large dark dots: seven high-confidence gamma-ray pulsars. Large light dots: three lower-confidence gamma-ray pulsars. Solid lines: timing age. Dotted line: open field line voltage. Dashed line: surface magnetic fields.}
\label{fig3}
\end{figure}

This sample of pulsars can be compared to other pulsars in terms some of the derived 
physical parameters.   Figure 3 is a distribution of pulsars as a function of their period 
and period derivative, with the gamma-ray pulsars shown as large dots.  Also shown 
are some of the derived physical parameters.  The gamma-ray pulsars tend to be 
concentrated (with the exception of the one millisecond candidate) in a region with 
high magnetic field (but not magnetar-strength) - shown by the dashed lines,  and 
relatively young ages - shown by the solid lines.  All ten gamma-ray pulsars share a 
third characteristic, shown by the dotted line, of having the open field line voltage 
high compared to most pulsars.  This is not surprising, since the particles are being 
accelerated electromagnetically.

\begin{figure}[b!] 
\centerline{\epsfig{file=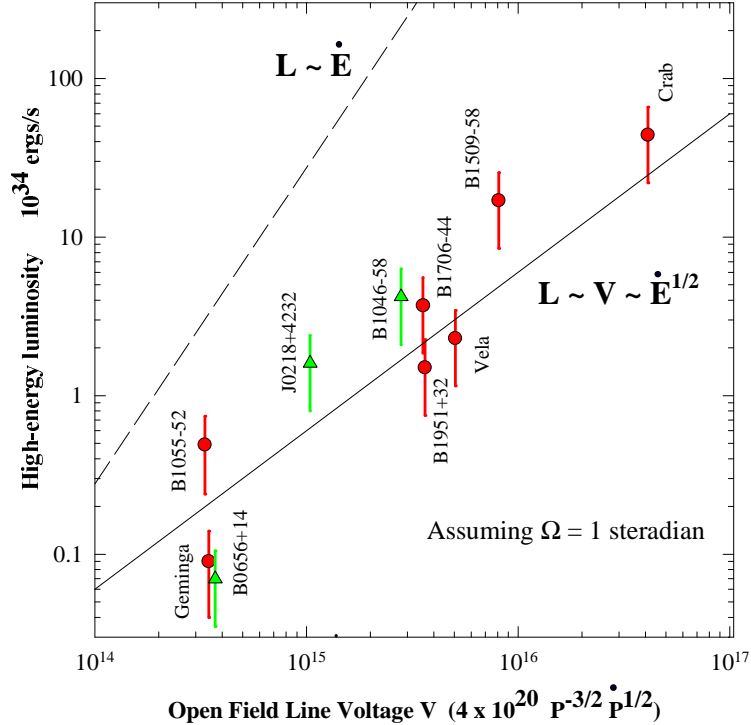,height=4.0in}}
\vspace{10pt}
\caption{High-energy pulsar luminosity as a function of the open field line voltage. Circles: high-confidence pulsars; Triangles: lower-confidence pulsars.}
\label{fig4}
\end{figure}

The open field line voltage, which is also proportional to the polar cap 
(Goldreich-Julian) current, has long been recognized as a significant parameter for 
gamma-ray pulsars [22].  Figure 5 illustrates that the pulsar high-energy luminosities, 
integrated above 1 eV, are approximately proportional to this parameter, shown by the 
solid line in the figure.  An interesting question is what happens for lower voltages, 
where the high-energy luminosity converges with the total spin-down energy 
available, shown by the dashed line.  This will be a question for future missions. 
\section*{Pulsars at the Highest Energies}

\begin{figure}[b!] 
\centerline{\epsfig{file=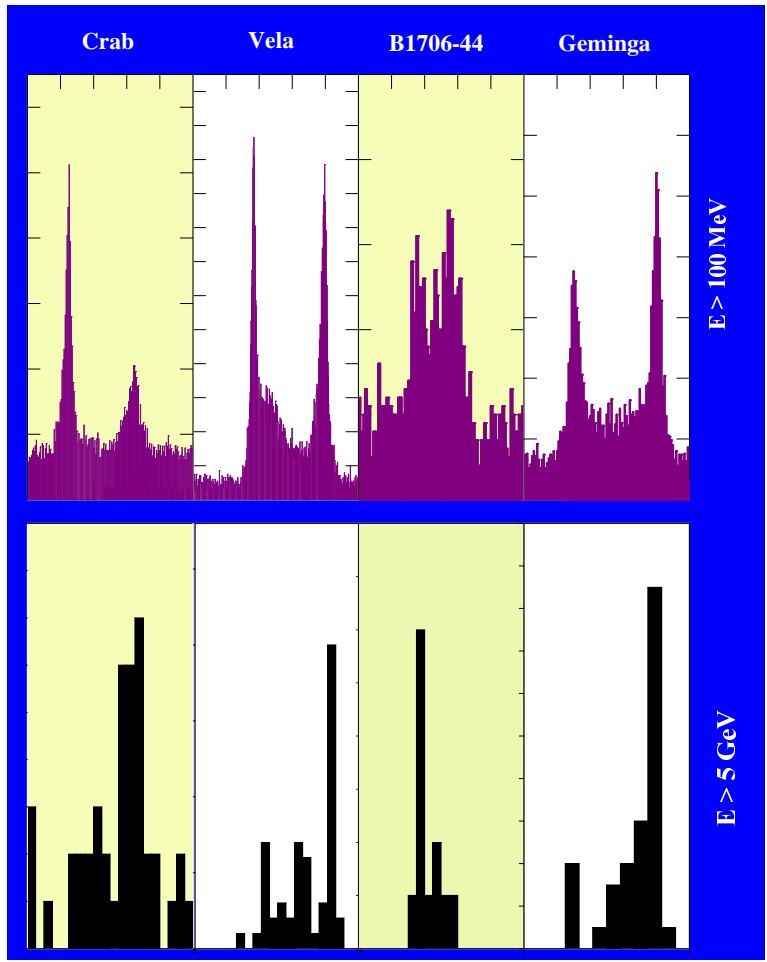,height=5.0in,bbllx=20pt,bblly=30pt,bburx=610pt,bbury=760pt,clip=.}}
\vspace{10pt}
\caption{Light curves of four gamma-ray pulsars above 100 MeV and above 5 GeV. Each panel shows one full rotation of the neutron star.}
\label{fig5}
\end{figure}

	What happens to pulsar observations at the highest energies?  No pulsars have 
been seen at TeV energies. The upper end of the EGRET range represents the highest 
energies for detections of pulsed emission.  There is evidence of pulsed emission 
above 5 GeV for all six of the pulsars definitely seen by EGRET.  Figure 5 shows the 
light curves for the four with the best statistics (the other two are consistent with 
these), shown in the upper panels in the 100 MeV range, and at the bottom for photons 
above 5 GeV.   It is clear that something has changed in the multi-GeV light curves - 
each of the pulsars above 5 GeV is dominated by one of the two pulses seen at lower 
energies.  In fact, for all except PSR B1706$-$44, it is the trailing pulse that dominates.  
Because of EGRET's very low background, the second pulse is still barely visible. A 
detector with higher background would see only the single pulse. These observations 
suggest that this energy range is a critical one for gamma-ray production.

\begin{figure}[b!] 
\centerline{\epsfig{file=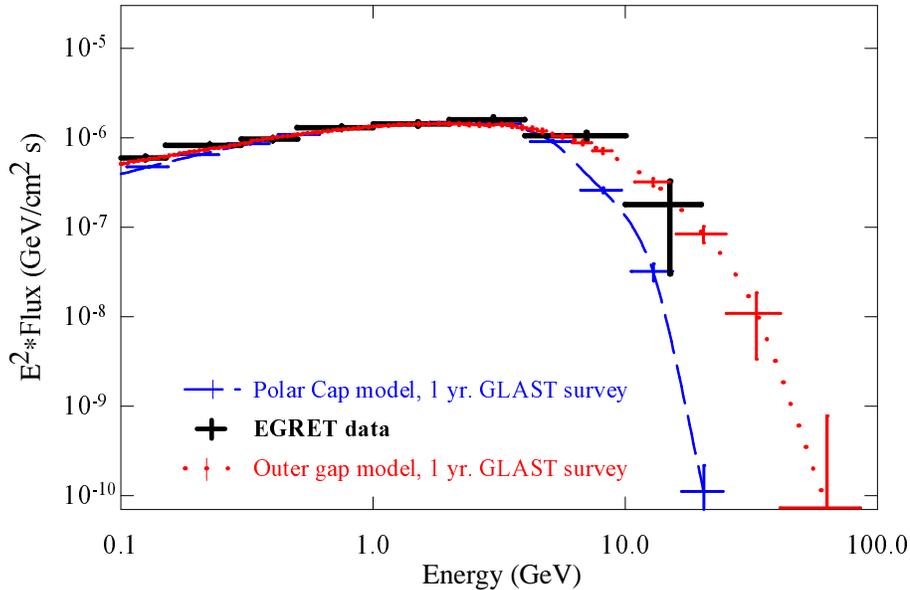,height=3.3in}}
\vspace{10pt}
\caption{High-energy spectrum of the Vela pulsar. Heavy error bars: EGRET data [27]. Dotted line: 
Outer gap model [25,26]. Dashed line: Polar cap model [23,24]. Error bars shown on the models are 
those expected from the GLAST mission [28] in a one-year sky survey.}
\label{fig6}
\end{figure}

The importance of the 1-20 GeV band is also visible in the energy spectra [17]. 
In all gamma-ray pulsars, the dominant power is seen in the hard x-ray to hard 
gamma-ray band - between 100 keV for the Crab and greater than 10 GeV for PSR 
B1951+32.   Also in all cases, there is a fall-off at higher energies.  The upper limits 
from the ground-based detectors are typically an order of magnitude or more below 
the peak luminosity. 
	The gap between about 10 GeV (where EGRET runs out of photons) and the 
current generation of ground-based telescopes is very important.  It is hard to predict 
from the limited EGRET data what is expected even at 100 GeV. Figure 6 shows the 
spectrum of Vela, the brightest of the EGRET pulsars. This spectrum is shown in $\nu$F$\nu$
or power per logarithmic energy interval format.  The 10-30 GeV point is based on 
only 4 photons, and the absolute calibration of EGRET at these energies is fairly 
uncertain. Also shown are the spectra expected in two popular gamma-ray pulsar 
models, the polar cap [23,24] and the outer gap [25,26].  The large error bars on the 
EGRET data make them consistent with both models.  The extrapolation to higher 
energies is, however, dramatically different.  There is also the possibility of a second, 
inverse-Compton component of the pulsed radiation at higher energies,  expected in 
some outer gap models [25], although there is yet no observational evidence for that  
component.  Searching for that second component will be an important goal of the 
next generation of very-high-energy gamma-ray telescopes.

\begin{figure}[b!] 
\centerline{\epsfig{file=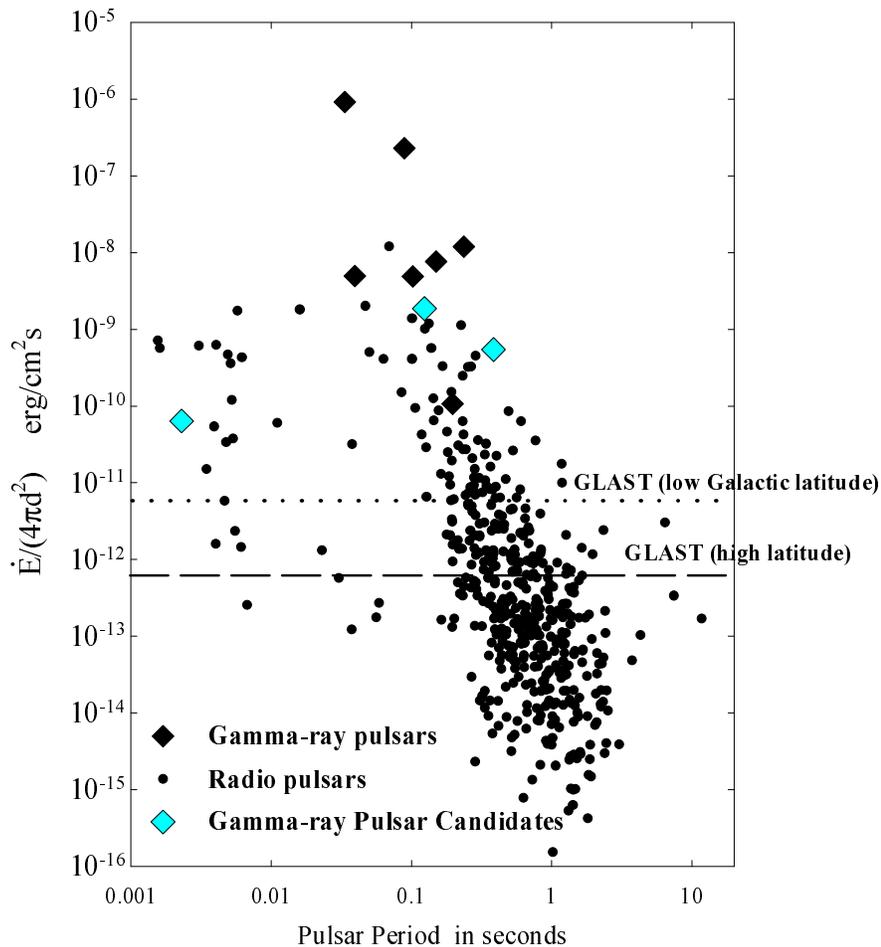,height=5.0in}}
\vspace{10pt}
\caption{Gamma-ray pulsar observability, as measured by the spin-down energy seen at earth.}
\label{fig7}
\end{figure}

\section*{Future Pulsar Observations at High Energies}

The high luminosity in the GeV energy range and the drop-off above that part 
of the spectrum suggest that GLAST will be the next big step in gamma-ray pulsar 
observations. AGILE will certainly make a contribution, especially in confirming 
those candidate pulsars for which the EGRET data are marginal, but GLAST will be 
needed for a major increase in sensitivity and energy reach.  Figure 6 shows one of the 
ways that GLAST will help - the smaller error bars on the theoretical curves show two 
models' predictions folded through the GLAST sensitivity for a one-year sky survey. 
GLAST will easily distinguish spectra such as these and might detect a second 
component at higher energies.
A second strength of GLAST will be in its much higher sensitivity than 
previous instruments.  Figure 7 shows one of the classic measures of pulsar 
observability: the pulsar spin-down luminosity divided by 4$\pi$ times the square of the 
distance, the total available pulsar energy output at Earth.  The 10 gamma-ray pulsars 
and candidates are shown as the large diamonds.  Six of the seven pulsars with the 
highest value of this parameter are gamma-ray pulsars. Below these, only a handful of 
pulsars are visible in gamma rays.  The GLAST sensitivity will push the lower limit 
down substantially farther.  Two sensitivity limits are shown for GLAST - one for 
low-Galactic-latitude sources and one for those at high latitudes, because the high 
diffuse emission along the Galactic plane reduces the sensitivity for point source 
detection.  The phase space that GLAST opens up is substantial.

\begin{figure}[b!] 
\centerline{\epsfig{file=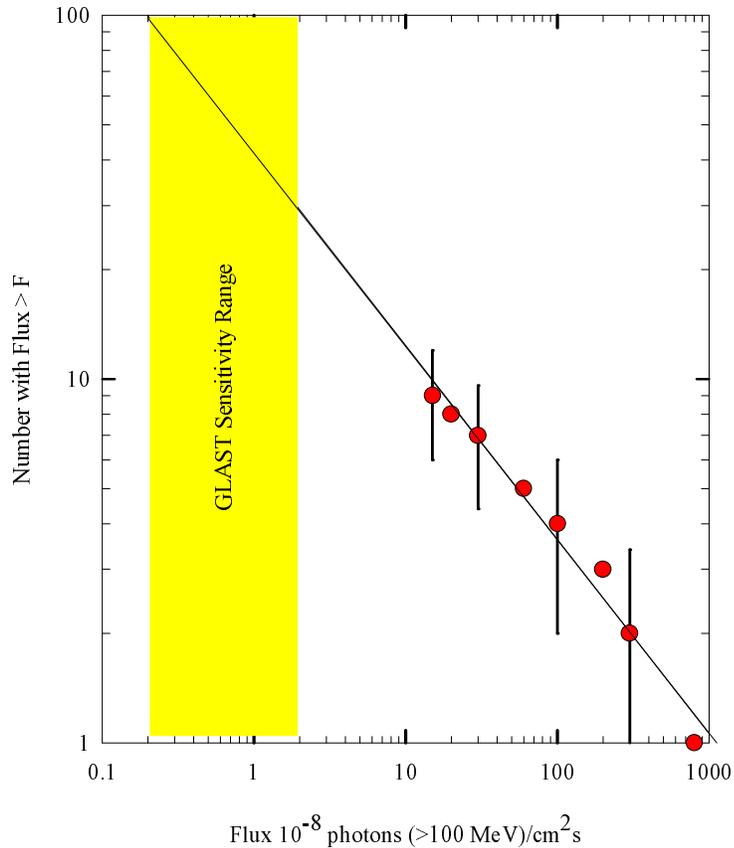,height=4.5in}}
\vspace{10pt}
\caption{Cumulative number of gamma-ray pulsars seen by EGRET as a function of 
observed flux.  Error bars are statistical and are shown on alternating data points for clarity.}
\label{fig8}
\end{figure}

How many pulsars will GLAST see?  To some extent that number depends on 
which model best describes the emission and how the pulsars are distributed on the 
sky.  One empirical estimate can be made by constructing a logN-logS curve from the 
known pulsars.   Figure 8 uses all nine of the high-energy gamma-ray pulsars to 
construct this function.  The simple linear fit suggests that GLAST might expect to 
detect between 30 and 100 gamma-ray pulsars.  The range of sensitivity is dependent 
on where the pulsars lie with respect to the Galactic plane.  In some respects this 
figure is pessimistic, possibly due to the small sample.  A disk population should 
increase linearly with decreasing sensitivity,  but the fitted line goes as the $-$0.5 power 
of the sensitivity.  A linear function would raise the number of pulsars to more than 
100.

\begin{table}
\caption{Predicted gamma-ray fluxes (units of 10$^{-8}$ ph [E$>$ 100 MeV] cm$^{-2}$s$^{-1}$)}
\label{table1}
\begin{tabular}{lrrrr}
Pulsar J &  Zhang \& Harding & Rudak \&Dyks & Romani  &  Cheng \&Zhang    \\
   &  [29]  &  [30]   &   [26]   &  [31]  \\
\hline
 1932+1059  & $<$580\tablenote{Calculated gamma-ray luminosity is a large fraction of the spin-down luminosity, hence an upper limit} & 90 & $\sim$0\tablenote{Observed gamma radiation expected to be small due to beaming} &  $<$16  \\
 0953+0755  & $<$500$^a$ & 90 & $\sim$0$^b$ &  $<$16  \\
 0437$-$4715  & $<$230$^a$ & $<$2 & $\sim$0 &  $<$16  \\
 2043+2740  & 50 & 30 & $\sim$0$^b$ &  50  \\
 1826$-$1334  & 20 & 20 & $\le$1 &  $<$16  \\
 1803$-$2137  & 20 & 20 & 30 &  $<$16  \\
 0742$-$2822  & 20 & 20 & $\le$7 &  25  \\
 0117+5914  & 20 & 20 & $\le$35 &  20  \\
 1801$-$2451  & 15 & 10 & 23 &  $<$16  \\
 1908+0734  & 50 & 10 & $\sim$0 &  $<$16  \\
 1730$-$3350  & 10 & 10 & $\le$20 &  $<$16  \\
 0538+2817  & 15 & 10 & $\le$24 &  20  \\
 0358+5413  & 10 & 8 & $\le$9 &  16  \\
 1453$-$6151  & 10 & $<$2 & $\sim$0$^b$ &  20  \\
 2337+6151  & 8 & $<$2 & $\le$11 &  20  \\
 1824$-$2452  & 16 & $<$2 & $\le$9 &  $<$16  \\
 1637$-$4553  & 4 & $<$2 & $\le$7 &  $<$16  \\
\end{tabular}
\end{table}

Just as interesting as the number of pulsars to be seen by GLAST is which 
particular pulsars will be seen.  Such observations are probably the best discriminator 
among models. Table 1 is a compilation with help from the theorists involved, 
showing the expected flux for various radio pulsars that have not yet been seen in 
gamma rays.  For many of these, different models make predictions that differ by 
more than an order of magnitude.  In these units, the GLAST sensitivity is 2 for 
pulsars near the plane and 0.2 for high-latitude pulsars; therefore GLAST will 
certainly provide solid observational tests of these and other gamma-ray pulsar 
models.

The third, and perhaps most important, capability of GLAST for pulsars will 
be its ability to find unknown pulse periods, in order to look for more Geminga-like 
pulsars.  Various analyses have shown that only the brightest pulsars could be found in 
the EGRET data without independent information [32-34]. For all of the other 
unidentified EGRET sources, the photons are just too far apart in time to derive 
unambiguous pulse periods.  With GLAST, periodicity searches will be feasible for 
most, if not all, the unidentified EGRET sources [35].  The potential is to find a whole 
new population of rotation-powered pulsars, much as x-ray astronomy has started to 
do in the past few years.  This could be an important new window onto the physics of 
the extreme conditions around these spinning neutron stars.

\section*{Summary}
\begin{itemize}
\item Gamma-ray pulsars remain a valuable probe of particle acceleration and 
interaction in the extreme conditions found near rotating neutron stars.

\item At least 7 pulsars are seen in gamma rays (six of those at energies above 100 MeV), with 3 additional good candidates.

\item The changing shape of the light curves and energy spectra in the 1-20 GeV range 
make this band particularly interesting for future observations.

\item GLAST, along with other satellite and ground-based gamma-ray telescopes, will 
make a major advance in gamma-ray pulsar studies.
\end{itemize}

\end{document}